%% file: Project.tex
\documentclass[a4paper,10pt]{article}
\usepackage[colorlinks,allcolors=DarkCyan,unicode,pdftitle={Localized oscillation of a Euler--Bernoulli beam}]{hyperref}

\usepackage[final]{graphicx}
\usepackage{eucal}
\usepackage{tempora}
\usepackage{amsmath}
\usepackage{amssymb}
\usepackage{amsthm}
\usepackage{graphicx} 
\usepackage{epsfig}
\usepackage{ifthen}
\usepackage{psfrag}
\usepackage[utf8]{inputenc}
\usepackage[english,russian]{babel}
\usepackage[T2A]{fontenc}
\usepackage{setspace}
\usepackage{mathtools}
\usepackage{url}
\usepackage[process=auto]{pstool}
\usepackage[svgnames]{xcolor}
\usepackage{stackrel}
\usepackage[sort&compress]{natbib}

\usepackage[left=2cm,right=2.5cm, top=2cm,bottom=3cm]{geometry}
\usepackage{authblk}

\input{style}

\input{def}

\title{Localized oscillations of an Euler--Bernoulli beam with time-varying
  parameters on a visco-elastic foundation:  asymptotics,  adiabatic
invariant, and equivalent Hamiltonian system}
\author{Ekaterina V. Shishkina}
\author{Serge N. Gavrilov}
\author{Yulia A. Mochalova}
\affil{Institute for Problems in Mechanical Engineering RAS, St.~Petersburg, Russia}

\begin{document}
\selectlanguage{english}
\maketitle

\begin{abstract} 
We consider localized oscillations of an Euler–Bernoulli beam on a
visco-elastic foundation coupled to a damped discrete oscillator. All system
parameters vary slowly and independently in time. For the conservative case,
we employ three different analytical approaches: an asymptotic method, a
technique based on the adiabatic invariance of the action of a trapped wave,
and an analysis of the equivalent Hamiltonian system. All three approaches
yield the identical formula for the oscillation amplitude. In the dissipative
case, the oscillation amplitude is obtained solely via the asymptotic
approach. Finally, we validate our analytical findings against numerical
simulations, demonstrating excellent agreement.
\end{abstract} 
\tableofcontents


\input{def-beam}

\input{fullbeam}

\end{document}

%% file: style.tex
\makeatletter

\@addtoreset{equation}{section}
\makeatother

%% file: def.tex
\def\TODO#1{\marginpar{\setstretch{1}\vspace*{0mm}\hrule\strut\vphantom{p}\scriptsize #1\strut\vphantom{p}\hrule}}

\def\d{\mathrm d}
\def\defe{\buildrel{\text{def}}\over=}
\def\tp{t^\prime}  
\def\prpr#1{#1^{\prime\prime}}
\def\thet#1{(\ref{#1})}
\def\veps{\varepsilon}
\def\Int{\int_{-\infty}^{+\infty}}
\def\IInt{\iint_{-\infty}^{+\infty}}
\def\VP{\operatorname {Vp}}
\def\PV{\operatorname {Vp}}
\def\const{\operatorname {const}}
\def\sign{\operatorname {sign}}
\def\Sqrt{\sqrt{\omega^2+k}}

\def\erf{\operatorname{erf}}
\def\S{{S}}
\def\C{{C}}
\def\defe{\buildrel{\text{def}}\over=}
\def\KK{{k_0}}
\def\Farg{\frac{\varkappa}{\sqrt{2\pi\mu}}}
\def\sign{\operatorname{sign}}
\def\const{\operatorname{const}}
\def\opm{{\omega^\pm}}
\def\Iint{\iint_{-\infty}^{+\infty}}  
\def\Np{\mathbf N^\prime}
\def\Rp{\mathbf R^\prime}
\def\bbU{\mathbf U}
\def\bbE{\mathbf E}
\def\bbF{\mathbf F}
\def\bbf{\mathbf f}
\def\bbN{\mathbf N}
\def\bbG{\mathbf G}
\def\bbL{\mathbf L}
\def\bbR{\mathbf R}
\def\bmR{\bmathcal R}
\def\bmF{\bmathcal F}
\def\mF{\mathcal F}
\def\bbr{\mathbf r}
\def\bbK{\mathbf K}
\def\bbi{\mathbf i}
\def\bbj{\mathbf j}
\def\ii{\bbi\,{\otimes}\,\bbi}
\def\dxi#1{{#1}_\xi}
\def\ddt#1{{#1}_{\tau\tau}}
\def\dt#1{{#1}_{\tau}}
\def\ddxit#1{{#1}_{\xi\tau}}
\def\ddxi#1{{#1}_{\xi\xi}}
\def\cD{\mathcal D}
\def\cK{\mathcal K}
\def\cEps{\mathcal E}
\def\pd#1#2{\dfrac{\partial#1}{\partial#2}}

\newcommand{\cc}{\mathrm{c.c.}}

\def\equ{equation\ }
\def\equs{equations\ }

\def\I{\mathrm i}

\def\res{\operatorname{Res}}

\definecolor{brown}{RGB}{163,40,40}

\def\i{\mathrm i}

\def\erf{\operatorname{erf}}

\def\Exp#1{\mathrm e^{#1}}

%% file: def-beam.tex
\def\defe{\buildrel{\text{def}}\over=}
\def\tp{t^\prime}  
\def\prpr#1{#1^{\prime\prime}}
\def\thet#1{(\ref{#1})}
\def\veps{\varepsilon}
\def\Int{\int_{-\infty}^{+\infty}}
\def\IInt{\iint_{-\infty}^{+\infty}}
\def\VP{\operatorname {Vp}}
\def\PV{\operatorname {Vp}}
\def\const{\operatorname {const}}
\def\sign{\operatorname {sign}}
\def\Sqrt{\sqrt{\omega^2+k}}
\def\erf{\operatorname{erf}}
\def\S{{S}}
\def\C{{C}}
\def\defe{\buildrel{\text{def}}\over=}
\def\KK{{k_0}}
\def\Farg{\frac{\varkappa}{\sqrt{2\pi\mu}}}
\def\sign{\operatorname{sign}}
\def\const{\operatorname{const}}
\def\opm{{\omega^\pm}}
\def\Iint{\iint_{-\infty}^{+\infty}}  
\def\Np{\mathbf T^\prime}
\def\Rp{\mathbf R^\prime}
\def\bbU{\mathbf U}
\def\bbE{\mathbf E}
\def\bbF{\mathbf F}
\def\bbf{\mathbf f}
\def\bbN{\mathbf T}
\def\bbG{\mathbf G}
\def\bbL{\mathbf L}
\def\bbR{\mathbf R}
\def\bmR{\boldsymbol{\mathscr R}}
\def\bmF{\boldsymbol{\mathscr F}}
\def\mF{\mathscr F}
\def\bbr{\mathbf r}
\def\bbK{\mathbf K}
\def\bbi{\mathbf i}
\def\bbj{\mathbf j}
\def\ii{\bbi\,{\otimes}\,\bbi}
\def\dxi#1{{#1}_\xi}
\def\ddt#1{{#1}_{\tau\tau}}
\def\dt#1{{#1}_{\tau}}
\def\ddxit#1{{#1}_{\xi\tau}}
\def\ddxi#1{{#1}_{\xi\xi}}
\def\cD{\mathscr D}
\def\cK{\mathscr K}
\def\cEps{\mathscr E}
\def\pd#1#2{\dfrac{\partial#1}{\partial#2}}
\def\pdd#1#2{\dfrac{\partial^2#1}{\partial#2^2}}

\def\w{{u}}

\def\equ{Eq.~}
\def\equs{Eqs.~}

\theoremstyle{remark}
\newtheorem*{Remark}{\it Remark}
\newtheorem{remark}{\it Remark}

\def\TODO#1{\marginparwidth=15mm
\marginpar{\hrule\strut\vphantom{p}\scriptsize #1\strut\vphantom{p}\hrule}}

\def\O{\Omega}
\def\TF{\mathcal T}
\renewcommand{\=}{\stackrel{\mbox{\scriptsize def}}{=}}
\newcommand{\mathcalC}{C}
\newcommand{\et}{{\epsilon t}}
\newcommand\pbaru{{\{\bar w\}}}
\newcommand\baru{{\bar w}}
\newcommand{\pup}{\{w\}}
\renewcommand\u{w}

%% file: fullbeam.tex
\section{Introduction}

The aim of this work is to develop approximate analytical methods for solving
problems concerning non-stationary strongly localized
\cite{luongo2001mode,Luongo1992}
or nearly localized oscillations in
discrete-continuous systems with time-varying parameters.
One notable, albeit not unique, example of such systems is found in moving
load problems \cite{Gavrilov2022jsv,gavrilov2002etm,Kruse1998,Gavrilov2024nody,Sergeev2005}, in which the load speed can be treated as a variable
parameter. The localized oscillations under consideration are associated with
a trapped mode 
\cite{ursell1951trapping,Ind-book-R2E,Mishuris2020,mciver2003excitation,Indeitsev2006,indeitsev2012motion,kaplunov1995simple,Glushkov2011a,Abramian1998,Abramyan1994}, 
which exists in the system. The corresponding mode frequency 
lies inside a stop-band of the dispersion relation.
By nearly localized oscillations, we mean a regime in which the motion of the
discrete subsystem asymptotically exhibits slowly decaying harmonic
oscillations due to either weak viscous damping
\cite{Abramian2025,Gavrilov2024nody,kaplunov1986torsional,Kruse1998,Roy2018} in the system or
energy radiation to infinity \cite{Lavrov2000}. The latter phenomenon is often observed in dynamic 
contact problems, in particular, in those involving moving loads \cite{gavrilov2004oscillation}.

Non-stationary oscillations in discrete-continuous systems with constant
parameters are frequently investigated by reducing the problem to an integral
equation governing the interaction force between the continuous and discrete
subsystems \cite{kaplunov1986torsional}. This approach is also typical for contact problems 
\cite{Lavrov2000,gavrilov2004oscillation,Johnson_KL--Contact_mechanics--1985}. In certain
cases, it can also be extended to systems with time-varying parameters \cite{Shishkina2019jsv,Gavrilov2022jsv,gavrilov2002etm,indeitsev2016evolution};
however the equation obtained can be investigated only numerically.
A
possible alternative is an asymptotic approach based on the assumption that the
system parameters vary sufficiently slowly \cite{abramyan2011oscillations,abramian2014oscillations}. For ordinary differential
equations (ODEs), this concept underlies the Liouville–Green approximation 
\cite{Liouville1837,Green1838,Green2014}
used to solve the governing equation of a harmonic oscillator with variable
stiffness. This approximation subsequently evolved into the WKB method
\cite{Jeffreys1925,Wentzel1926,Kramers_1926,Brillouin1926}, which
is widely employed in quantum mechanical problems. The WKB method can be
considered as a particular case of Nayfeh's method of multiple scales
\cite{nayfehperturbation}. A natural extension of both approaches to partial
differential equations (PDEs) describing non-stationary processes is the space-time
ray method \cite{Babich2009,Babich2002,Babich2025,Hinch2002}.

In \cite{gavrilov2002etm}, an asymptotic approach in the spirit of Nayfeh's
method of multiple scales was proposed to solve the problem of transverse
vibrations of an inertial load moving non-uniformly along a string on the
Winkler foundation. To describe the evolution of a trapped wave co-moving
with the load, the authors introduced an asymptotic ansatz that generalized
the ansatz employed in the space-time ray method for traveling waves. This
approach yielded an approximate analytical solution in a simple form that
demonstrated excellent agreement with numerical solutions of the corresponding
integral equation. Subsequently, a similar technique was employed in
\cite{Gavrilov2024nody} to address a substantially more challenging problem:
constructing an asymptotic solution for localized oscillations in a string on
the
Winkler foundation traversed by a mass-spring inclusion.
It was assumed that all parameters of the discrete-continuous system were
time-varying.
It
was found that the amplitude of the localized oscillations is a function of
the system parameters calculated for the current time, being independent of
their history. In addition, an analogous system was examined in which both
subsystems possessed weak damping, with the damping coefficients also treated
as time-varying. While an asymptotic solution was obtained in this case as
well, the amplitude could only be determined as a functional depending on the
past history of the system parameters.

A recent study \cite{arXiv.2602.18815} demonstrated that the ability to
determine the amplitude of localized oscillations as a function of the
instantaneous system parameters can be attributed to the adiabatic invariance
of the action of a trapped wave. The concept of the action of a trapped wave
generalizes both the classic action well-known from
Hamiltonian mechanics 
\cite{Arnold2009,Henrard1993} 
and the wave action known from the Whitham theory
\cite{Whitham1999,Buehler2009,Bretherton1966,Whitham1965,Whitham1967,Lighthill1967full,Babich2002}.
Although the asymptotic approach remains the only known framework capable of
rigorously justifying the invariance of the localized wave action for the
discrete-continuous system under consideration, postulating this action
invariance provides a technically far simpler route to determining the
amplitude evolution of localized oscillations. Implementing the asymptotic
approach requires one to guess the structure of the asymptotic ansatz a
priori, which can be challenging. However, the method based on adiabatic
invariants also entails its own pitfalls, stemming from the ambiguity in
choosing the energy functional.

Finally, the same work \cite{arXiv.2602.18815} proposed an approach that
simplifies the solution of the problem even further. It turns out that, at
least for the specific problem treated in \cite{arXiv.2602.18815}, the
non-dissipative discrete-continuous system under consideration can be associated with
an ``equivalent'' Hamiltonian system possessing the same adiabatic invariant.
In this case, the parameters of the equivalent system can be determined by
analyzing the original problem in the leading-order approximation, i.e., the
problem with constant parameters. 

In this paper we apply the approaches recently developed in \cite{Gavrilov2024nody,arXiv.2602.18815}
to the problem concerning localized oscillation of an Euler–Bernoulli beam on
a visco-elastic foundation coupled to a damped discrete oscillator. All parameters of the system independently vary in time in a slow manner. The
problem under consideration was previously addressed in \cite{Shishkina2019jsv}
in a very particular case. Namely, it was assumed:
\begin{itemize} 
  \item
  The inclusion was not a discrete mass-spring system, but a discrete spring with a negative stiffness.
 \item 
   The only time-varying parameter of the system was the stiffness of the discrete spring.
 \item
   Only the non-dissipative case was under consideration.
\end{itemize} 
And finally, only the asymptotic approach was used in \cite{Shishkina2019jsv},
whereas two additional analytic techniques, namely, the method based on the adiabatic invariance
of the action of a trapped wave, and the consideration of the equivalent Hamiltonian system, are 
utilized in the current paper in the non-dissipative case. These approaches
are significantly simpler than the asymptotic one. We show that these two 
approaches are applicable not only in the case of a string considered in 
\cite{Shishkina2019jsv}, but also in the case of an Euler--Bernoulli beam.

The paper is organized as follows. 
In Sect.~\ref{sect-formulation}, we provide the mathematical formulation for
the problem under consideration. 
In Sect.~\ref{sect-anzatz}, we treat the problem by means of asymptotic
approach. 
In Sect.~\ref{sect-invariant}, we use the method based on the adiabatic
invariance of the action of a trapped wave. 
In Sect.~\ref{sect-ham}, we use the approach based on considering the equivalent
Hamiltonian system.
In Sect.~\ref{sect-num}, we verify analytic results by numerical calculations.
In the Conclusion (Sect.~\ref{sect-concl}), we discuss the basic results of the paper
and possible generalizations. In the Appendix, we provide basic formulae describing
solutions of the 
corresponding zeroth-order problem for the equation with constant
coefficients.

\section{Mathematical formulation}
\label{sect-formulation}
The transverse oscillation of an Euler--Bernoulli beam on the Winkler
foundation with an embedded discrete oscillator is governed by the following
equation:
\begin{equation}
(\rho 	\dot w)^{\boldsymbol\cdot} +B w''''
  + kw +2 \epsilon \gamma \dot w
 = \Big( 
 p(t) - \big(M	\dot w \big)^{\boldsymbol\cdot} - K w - 2\epsilon \varGamma \dot w 
\Big)\delta(x).
	\label{eq}
\end{equation}
Here $w(x, t)$ is the displacement, $B$ is  the flexural
stiffness, $\rho$ is the mass density of the beam, $\epsilon \gamma$ is the damping in the foundation,
$M$ is the discrete oscillator mass,
$K$ is the oscillator stiffness,  
$k$ is the Winkler foundation stiffness , 
$\epsilon\varGamma$ is the damping in the oscillator,
$x$ is the spatial coordinate, $t$ is time,  $\epsilon$ is
a formal small parameter,
$\delta(\cdot)$ is the Dirac delta-function. Prime and overdot denote the partial derivatives
with respect to $x$ and $t$, respectively.
All the system parameters are smooth functions of the slow time $\epsilon t$:
\begin{equation}
\begin{gathered}	
K=K(\epsilon t),
\qquad
M=M(\epsilon t),
\\
B=B(\epsilon t),
\quad
\rho=\rho(\epsilon t),
\quad
k=k(\epsilon t),
\\
\varGamma=\varGamma(\epsilon t),
\quad
\gamma=\gamma(\epsilon t).
\end{gathered}
\label{all-slow-et}
\end{equation}

We assume that the external loading $p$ is a pulse of a finite
duration:
\begin{equation}
p({t})\equiv 0, \qquad {t}<0 \quad\text{or}\quad {t}>{t}_0>0
\label{pulse}
\end{equation}
for some $t_0>0$. In what follows, we call the particular case of the system under consideration
where 
\begin{equation}
 \varGamma=0,\qquad\gamma=0
 \label{non-diss}
\end{equation}
the non-dissipative case. If 
 Eq.~\eqref{non-diss} is not fulfilled, we deal with the dissipative case.

The problem can be formulated as the following equivalent one, which consists of the governing equation~\eqref{eq} with zero right-hand side for $x \neq \pm 0$: 
\begin{equation}
(\rho 	\dot w)^{\boldsymbol\cdot} +B w''''
  + kw +2 \epsilon \gamma \dot w
  =0
	\label{eq-semi}
\end{equation}
and the four boundary conditions at $x=0$:
\begin{gather}
	[w]=0,
	\\
	[w']=0,
	\\
	[w'']=0,
	\\
	B[w''']= 
 p(t) - \Big(\big(M	\dot w \big)^{\boldsymbol\cdot} + K w + 2\epsilon \varGamma \dot w 
 \Big)\Big|_{x=0}.
	\label{bc3}
\end{gather}
Here, $[\mu]\equiv\mu(x+0)-\mu(x-0)$ for any arbitrary quantity
$\mu$. The initial conditions can be formulated in the following form \cite{Vladimirov1971}:
\begin{equation}
 {\u}\big|_{t<0}\equiv0.
 \label{ic}
\end{equation}
We are interested in localized solutions with finite energy. Accordingly, we
can require 
\begin{equation}
2E\pup<\infty
\label{pup}
\end{equation}
where 
\begin{gather}
2E\pup=2E_{\mathrm c}\pup+2E_{\mathrm d}\pup,
	\label{energy-unpert}
\end{gather}
is the total energy of the beam:
\begin{gather}
  E_{\mathrm c}\pup=\frac14\int_{-\infty}^{+\infty} 
  \big(
    \rho\dot\u^2 + B(\u'')^2  + {k {\u}^2} 
\big )\, \d x,
	\label{energy-unpert-c0}
  \\
  E_{\mathrm d}\pup=\frac14\big(M\dot\u^2 + 	K {\u}^2\big) \big\vert_{x=0}.
	\label{energy-unpert-d}
\end{gather}
Condition 
\eqref{pup} plays the role of a boundary condition at infinity.

The formulated problem \eqref{eq-semi}--\eqref{energy-unpert-d}
is symmetric with respect to $x=0$, therefore, it is useful to reformulate it
for a semi-infinite beam $x>0$. In what follows, we consider 
the governing equation~\eqref{eq-semi} for $x>0$, subjected to the two boundary conditions
at $x=0$:
\begin{gather}
	w'=0,
	\label{angle-half-beam-dim}
	\\
	2Bw'''=
 p(t) - \Big(\big(M	\dot w \big)^{\boldsymbol\cdot} + K w + 2\epsilon \varGamma \dot w 
 \Big)\Big|_{x=0}
	\label{bc3-semi}
\end{gather}
initial condition \eqref{ic}, and condition 
\eqref{pup} where 
\begin{gather}
  E_{\mathrm c}\pup=\frac12\int_0^{+\infty} 
  \big(
    \rho\dot\u^2 + B(\u'')^2  + {k {\u}^2} 
\big )\, \d x.
	\label{energy-unpert-c}
\end{gather}

\section{Asymptotics}
\label{sect-asy}

\subsection{Ansatz}
\label{sect-anzatz}

We look for the solution under the following assumptions:
\begin{enumerate}
	\item $\epsilon=o(1)$,
	\item $t=O(\epsilon^{-1})$,
  \item The localization condition~\eqref{K-restr} is satisfied for all $t$.
\end{enumerate}
We also require that the frequency equation for the trapped mode and the dispersion relation at $x\geq0$ hold for all $t$.
We use the following asymptotic ansatz:
\begin{gather}
	w(x,t)=H(T)\sum_{\alpha=\pm1} W^\alpha(X,T)\,\exp\phi^\alpha+\cc,
	\label{FB-slow-repr-BEAM-new-dim}
  \\
	X=\epsilon x, \quad T=\epsilon t,
	\\
  ({\phi^\alpha})'=\I\omega^\alpha(X,T),
  \qquad
  {\dot\phi^\alpha}{}=-\I\Omega^\alpha(X,T),
	\label{FB-fast-phases-BEAM-new-dim}
	\\
	\phi^\alpha(0,0)=- \I \frac{\alpha\phi_0}{2}, 
	\\
  \phi_0=\frac\pi2,
  \label{phi0pi2}
  \\
	\Omega^\alpha(0,T)=\Omega_0(T),
	\label{Omega(0,T)}
	\\
	W^\alpha(X,T)=\sum_{j=0}^{\infty} \epsilon^{j}W^\alpha_j(X,T).
	\label{solution-series-new-dim}
\end{gather}
Here the superscript $\alpha=\pm 1$, $\Omega_0$ is the root of the frequency equation~\eqref{fr-eq-dim},
\begin{equation}
  \Omega_\ast=\sqrt{\frac{k}{\rho}}
	\label{w-a-half-beam-dim}
\end{equation}
is the cut-off frequency,
$\omega^\alpha$ is the root 
\eqref{omega-delta}
of the dispersion relation~\eqref{disp} calculated for $\Omega=\Omega_0$, 
$H(\cdot)$ is the Heaviside step-function,
symbol $\cc$ denotes complex
conjugate terms.
We also require that the wave-number $\omega^\alpha(X,T)$ and the frequency $\Omega^\alpha(X,T)$ satisfy
the eikonal equations
\begin{equation}
  (\Omega^\alpha)'_X+(\omega^\alpha)'_T=0,
	\label{FB-dxx-BEAM-dim}
\end{equation}
which express the fact that
\begin{equation}
  (\dot\phi^\alpha)'=(\phi^\alpha{}')^{\boldsymbol\cdot}.
\end{equation}
Thus,
\begin{equation}
	\phi^{\alpha}(x,t)=\I\int_{(0,0)}^{(x,t)}
  \big( \omega^{\alpha}(\epsilon \hat{x},\epsilon \hat {t})\,\d\hat{x}-
    \Omega^{\alpha}(\epsilon \hat{x},\epsilon \hat{t}) \, \d\hat{t}\,\big)
  - \I\frac{\alpha\phi_0}{2}.
	\label{phase_int}
\end{equation}
Slow variables $X$, $T$ and two fast phases $\phi^\alpha$ are assumed to be the new independent variables. 
We additionally assume that
\begin{equation}
	W_0^{1}(0,T)=W_0^{-1}(0,T).
	\label{W-eq-dim}
\end{equation}
We emphasize that generally 
\begin{equation}
  W_0^{1}{}'_X(0,T)\neq W_0^{-1}{}'_X(0,T),\qquad
  \Omega_0^1{}'_X(0,T)\neq \Omega_0^{-1}{}'_X(0,T).
\end{equation}

\begin{remark} 
In the zeroth order approximation $\epsilon=0$, the asymptotic ansatz 
\eqref{FB-slow-repr-BEAM-new-dim}--\eqref{W-eq-dim} transforms into the form
of free localized oscillations described by Eq.~\eqref{free-osc}.
\end{remark} 

\subsection{Derivation and solving of the first approximation equation}
\label{sect-asolve}

The differential operators can be represented in the following form:
\begin{gather} 
	\partial_t=-
	\I\,\Omega_0\,\partial_{\phi }+\epsilon\,\partial_{T},
	\label{FB-BEAM:'T:1-new-dim}\\
	\partial_{tt}^2=-\Omega_0^2\,\partial^2_{\phi \phi }
	-2\epsilon \I\,\Omega_0\,\partial^2_{\phi T}
	-\epsilon \I\, {\Omega_0'}_T\,\partial_{\phi }
	+O(\epsilon^2),
	\label{FB-BEAM:''T-new-dim}\\
	\partial_{x}=
	\I\,\omega\,\partial_{\phi}+\epsilon\,\partial_{X},
	\label{FB-Bpdx1-new-dim}
	\\
	\partial^2_{x x}=-\omega^2\,\partial^2_{\phi\phi}
	+\epsilon (2\I\,\omega\,\partial^2_{\phi X}
	+\I\, {\omega}'_X\,\partial_{\phi})
	+O(\epsilon^2),
	\label{FB-Bpdx2-new-dim}
	\\
	\partial^3_{x x x} = -i\omega^3\partial^3_{\phi\phi\phi}+
	\epsilon(-3 \omega ^{2} \partial^3_{\phi\phi X}   
	-3  \omega \omega'_X \partial^2_{\phi\phi})
	+O(\epsilon^2),
	\label{FB-Bpdx3-new-dim}
	\\
	\partial^4_{x x x x} = 
	\omega^4\partial^4_{\phi\phi\phi\phi}+
	\epsilon(-4\I \omega ^{3} \partial^4_{\phi\phi\phi X}   
	-6 \I  \omega    ^{2} \omega'_X \partial^3_{\phi\phi\phi})
	+O(\epsilon^2).
	\label{FB-Bpdx4-new-dim}
\end{gather}
Here and in what follows, for the aim of simplicity, we skip the superscript
$\alpha$ for $W^\alpha$,  $\phi^\alpha$, $\Omega^\alpha$, $\omega^\alpha$ in
expressions, which are valid for any $\alpha$.

Substituting the ansatz~\eqref{FB-slow-repr-BEAM-new-dim}-\eqref{solution-series-new-dim} 
into the governing equation~\eqref{eq-semi} at $x \to 0$ and equating the coefficients of like powers $\epsilon$, one obtains that to the first-order approximation:
\begin{equation}
  4B\omega^3{W_0}{}'_X + 6B\omega^2\omega{}'_X W_0
  + \rho\big(2\Omega_0 {W_0}{}'_T + \Omega_0{}'_T W_0 \big)+\Omega_0 {\rho}'_T W_0+2\gamma\Omega_0 W_0= 0.
	\label{gov-eq-half-beam-mc-1app-dim}
\end{equation} 
The corresponding first-order approximations of boundary conditions~\eqref{angle-half-beam-dim},
\eqref{bc3-semi} are
\begin{gather}
  \sum_{\alpha=\pm1} \Exp{-\I  \frac{\alpha\pi}{4}}{W_0}{}'_X=0,
	\label{angle-half-beam-1app-dim}
\\
\begin{multlined}
  -3B\sum_{\alpha=\pm1}\Exp{-\I  \frac{\alpha\pi}{4}}
  \big(\omega^2{W_0}{}'_X + \omega \omega{}'_X W_0 \big)
	\\
  \qquad\qquad
  \qquad\qquad
  =\I \sum_{\alpha=\pm1}\Exp{-\I \frac{\alpha \pi}{4}} \left (M\Omega_0 {W_0}{}'_T+\frac{M}{2}{\Omega_0}'_T W_0 + \frac{M'_T}{2}  \Omega_0  W_0  +
	\Omega_0\varGamma W_0  \right).
	\label{force-half-beam-1app-dim}
\end{multlined}
\end{gather}

One has
\begin{multline}
	\omega'_X=\omega'_\Omega \Omega'_X = -\omega'_\Omega \omega'_T
  =
  -\omega'_\Omega(\Omega_0{}'_T+\omega'_B B'_T+\omega'_\rho \rho'_T+\omega'_k k'_T)\\=
  \frac{(\alpha+\I)^2}{16B\sqrt{\frac{B}\rho} \sqrt{\frac k\rho-\Omega_0^2}}
  \left(
    -\Omega_0 (-k+\rho\Omega_0^2)B'_T
    -\Omega_0 k'_T
    +\Omega_0^3\rho'_T
    +\rho\Omega_0^2\Omega_0{}'_T
  \right)
	\label{FB-dxx-BEAM-conseq-dim}
\end{multline}
due to dispersion relation~\eqref{disp} and the eikonal equation~\eqref{FB-dxx-BEAM-dim}.
By virtue of Eq.~\eqref{gov-eq-half-beam-mc-1app-dim} and \eqref{FB-dxx-BEAM-conseq-dim}
one can demonstrate that 
\begin{gather}
  {W_0}{}'_X\big|_{\alpha=\pm1}=(\I+\alpha) F,
	\label{W_x_gen-half-beam-dim}
  \\
  \begin{multlined} 
    F=\Lambda^{-1}
  \Big(3\rho\Omega_0(\rho\Omega_0^2-k)W_0B'_T
  +B\Omega_0\left(3\rho k'_TW_0
  +8(\rho\Omega_0^2-k)(\gamma W_0 +\rho W_0{}'_T)
  +(\rho\Omega_0^2-4k)W_0\rho'_T\right)
\\
  -2B\rho(2k+\rho\Omega_0^2)W_0\Omega_0{}'_T
  \Big),
  \end{multlined} 
\\
 \Lambda= {16\sqrt2\left(\frac{B}\rho\right)^{5/4}\rho^2\left(\frac k\rho-\Omega_0^2\right)^{3/4}(\rho\Omega_0^2-k)}.
\end{gather}
Thus, Eq.~\eqref{angle-half-beam-1app-dim} 
can be rewritten as
\begin{equation}
  \Exp{-\frac{\I\pi}4}(\I+1)
  +
  \Exp{\frac{\I\pi}4}(\I-1)=0,
\end{equation}
which is equivalent to
\begin{equation}
\tan\frac{\pi}4=1,
\end{equation}
i.e., 
Eq.~\eqref{angle-half-beam-1app-dim} is fulfilled.

Taking into account Eqs.~\eqref{w-a-half-beam-dim},
\eqref{gov-eq-half-beam-mc-1app-dim}, \eqref{FB-dxx-BEAM-conseq-dim}
we can transform Eq.~\eqref{force-half-beam-1app-dim} into the form of the following 
ordinary differential equation for $W_0$:
\begin{multline}
  \frac{W_0{}'_T}{W_0}=-\frac{1}{G}\Bigg( 3 \sqrt[4]{B}\gamma+\sqrt{2}\varGamma\sqrt[4]{k-\rho \Omega_0^2}+\frac{3\rho B'_T}{8\sqrt[4]{B^3}}-\frac{3\sqrt[4]{B}\rho k'_T}{8(k-\rho \Omega_0^2)}+\frac{\sqrt{2}}{2}\sqrt[4]{k-\rho\Omega_0^2}M'_T
	\\
	+\frac{3\sqrt[4]{B}(4k-3\rho\Omega_0^2)\rho'_T}{8(k-\rho\Omega_0^2)}
	+\frac{\left(3\sqrt[4]{B}\rho (2k-\rho \Omega_0^2)+2\sqrt{2}M\sqrt[4]{(k-\rho\Omega_0^2)^5}\right){\Omega_0}'_T}{4\Omega_0(k-\rho\Omega_0^2)}
	\Bigg),
	\label{od-eq-dim}
\end{multline}
\begin{equation}
	G=3\sqrt[4]{B}\rho + \sqrt{2}M\sqrt[4]{k-\rho\Omega_0^2}.
		\label{od-eq-G-dim}
\end{equation}
To derive the last two equations Wolfram Mathematica software was used.

One can verify that the general solution of Eq.~\eqref{od-eq-dim} can be represented in the form:
\begin{equation}
  W_0={\mathcal A} \mathcal C_0(T)\exp{\big(-D(T)\big)},
	\label{sol-dim}
\end{equation}
where
\begin{equation}
	\mathcal C_0(T)=\frac{\sqrt[4] 2\sqrt[8]{k-\rho\Omega_0^2}}{\sqrt{\Omega_0 \big( 3\rho\sqrt[4]{B} + \sqrt{2}M\sqrt[4]{k-\rho\Omega_0^2} \big)}},
  \label{amplitude}
\end{equation}
\begin{equation}
	D(T)=\int_{0}^{T}\frac{3\sqrt[4]{B}\gamma + \sqrt{2}\varGamma\sqrt[4]{k-\rho\Omega_0^2}}{3\sqrt[4]{B}\rho + \sqrt{2}M\sqrt[4]{k-\rho\Omega_0^2}} \,d\hat{T}.
	\label{sol-dim-D}
\end{equation}
Here ${\mathcal A}$ is an arbitrary complex constant. 

Due to Eqs.~\eqref{phi0pi2}, \eqref{Omega(0,T)}, \eqref{phase_int},
\begin{equation}
	\phi^{\alpha}(0,t)=-\I\left(\int_0^t \Omega_0(\epsilon \hat{t}) \, \d\hat{t} + \frac{\alpha\pi}{4}\right).
\end{equation}
Now according to Eqs.~\eqref{FB-slow-repr-BEAM-new-dim}--\eqref{W-eq-dim}, one gets
\begin{equation}
  w(0,t)=4|{\mathcal A}|H(T)\mathcal C_0(T)\exp{\big(-D(T)\big)}\cos\left( \frac{\pi}{4}\right)\cos(\psi-\arg \mathcal A)+O(\epsilon),
	\label{eval-osc-form}
\end{equation}
where 
\begin{gather} 
\psi \equiv \int_0^t \Omega_0(\epsilon\hat{t}) \, \d\hat{t}.
\label{psi-def}
\end{gather}

For $T \to +0$, the leading-order term of Eq.~\eqref{eval-osc-form} should
coincide with the leading-order term of
Eq.~\eqref{undamped-as-general-pnotzero-dim}. 
The detailed discussion concerning the matching procedure can be found in 
\cite{arXiv.2602.18815,Gavrilov2024nody}.
For the problem under consideration, the matched solution 
can be represented in the following form:
\begin{gather}
w(0,t) =
|\mathcal C(\epsilon t)| \exp{\big(-D(\et)\big)} 
\cos\big(\psi - \arg \mathcal C\big)+\text{o}(1), 
\label{sol-eg0}
\\
|\mathcal C|=|\mathcal C(\epsilon t)|=|C(0)|{\frac{|\mathcal C_0(\epsilon t)|}{|\mathcal C_0(0)|}},
\label{amp}
\\
\arg \mathcal C=\arg C(0),\qquad
\arg \mathcal C_0=\arg C_0(0),
\label{phase}
\end{gather}
where $|C_0(0)|$, $|C(0)|$, $\arg C(0)$, and $\arg C_0(0)$ are defined by Eqs.~\eqref{C00-mod}--\eqref{Co-value-eff}. 
Note that 
\begin{equation}
\mathcal C(\et)=\mathcal C_0(\et),\qquad
 C(0)=C_0(0)
\end{equation}
in the case $p=\delta(t)$. Quantities $\mathcal C(\et),\ \mathcal C_0(\et)$ are amplitudes
in the system with time-varying parameters 
\eqref{all-slow-et}; quantities $C(0),\ C_0(0)$ 
are amplitudes
in the system with constant parameters 
\eqref{all-slow0}.

In the non-dissipative case \eqref{non-diss},
similarly to the string problem \cite{Gavrilov2024nody}, the amplitude of localized oscillations is obtained as a function of the instantaneous values of the system parameters.
This result is associated with the fact that the right-hand side
of Eq.~\eqref{od-eq-dim} is the exact derivative of a certain function of
the system parameters,
provided that Eq.~\eqref{non-diss} is fulfilled. In the dissipative case,
generally speaking, the amplitude is a functional of the history of the system
parameters. However, in some exceptional cases
the amplitude can be represented as a function of the current parameter values and the current
time.
We can underline at least the following such cases:
\begin{enumerate}       
\item
The limiting case 
of an isolated mass-spring-damper system with
$\varGamma/M=\const$, where $K$ is a variable quantity, 
 considered in detail in
\cite{Gavrilov2024nody} (Appendix~A);
\item Limiting case $\rho=0,\ \gamma=0$
with
$\varGamma/M=\const$,
where $K$, $k$, $B$ are variable quantities;
\item The case $M=0$, $K=\const$, $\varGamma=\const\neq0$, 
$B=\const$, $\rho=\const$, $\gamma=0$, where $k$ is a variable
quantity;
\item The case  $M=0$, $\varGamma=0$, $\gamma/\rho=\const$, where
$K$, $B$, $k$ are variable quantities.
\end{enumerate}
To prove this proposition in cases 3 and 4,
analogously to 
\cite{Gavrilov2024nody}, we need to use 
the frequency equation 
\eqref{fr-eq-dim}.

\begin{remark} 
  In the particular case 
  \begin{equation}
    \rho=1,\qquad B=1,\qquad k=1, \qquad M=0,
  \end{equation}
 according to Eq.~\eqref{amplitude},
 we get the following formula:
 \begin{equation}
   \mathcal C\propto\frac{\sqrt[8]{1-\Omega_0^2}}{\sqrt\Omega_0}
 \end{equation}
that is in agreement with the main result of the previous paper \cite{Shishkina2019jsv}.
\end{remark} 

\section{Adiabatic invariant}
\label{sect-invariant}
In the rest of the paper, we deal with the non-dissipative case \eqref{non-diss}.


Following \cite{arXiv.2602.18815},
let us show that the formula describing the evolution of the amplitude for the
trapped wave can be obtained by postulating that the action of the trapped
wave
\begin{equation}
  J=\left.
  \frac {E\{\bar \u\}}{\Omega_0}
    \right|_{C=\mathcal C}
\end{equation}
is an adiabatic
invariant.
Here, $E$ denotes the total energy functional defined by Eqs.~\eqref{energy-unpert},
	\eqref{energy-unpert-c},
	\eqref{energy-unpert-d},
and $\baru$ is defined by the right-hand side of Eq.~\eqref{free-osc}.
Substituting Eq.~\eqref{free-osc}
into Eqs.~\eqref{energy-unpert-c}, \eqref{energy-unpert-d}, respectively, one gets
\begin{gather}
  \frac{2 E_{\mathrm c}}{{C}^2}
  =\frac{3\rho\Omega_0^2\sin^2 \psi_0}{8a(\Omega_0)}
  +
  \frac{Ba^3(\Omega_0)\cos^2\psi_0}{2}
  +
\frac {3k\cos^2\psi_0}{8a(\Omega_0)},
\label{Ec}
\\
\frac{2E_{\mathrm d}}{{C}^2}=\frac{M\Omega_0^2\sin^2\psi_0}4+ \frac{K\cos^2\psi_0}4,
\label{Ed}
\end{gather}
where 
\begin{equation}
 \psi_0
 =\Omega_0t-\arg C.
\end{equation}
Substituting Eqs.~\eqref{a-def}, \eqref{Ec}, \eqref{Ed} into Eq.~\eqref{energy-unpert} results in:
\begin{multline} 
  \frac{2 E_{\mathrm c}}{\mathcalC^2}
  +
  \frac{2 E_{\mathrm d}}{\mathcalC^2}
  =
  \frac{B^{1/4}}{8\sqrt2(k-\rho\Omega^2)^{1/4}}
  \left(
    2\rho\Omega_0^2
    +
    \sqrt2(M\Omega_0^2+K)\left(\frac{k-\rho \Omega_0^2}{B}\right)^{1/4}+8k\cos^2\psi_0
  \right.
  \\+
  \left(-4\rho\Omega_0^2+\sqrt2(K-M\Omega_0^2)\left(\frac{k-\rho \Omega_0^2}{B}\right)^{1/4}\right)
\cos2\psi_0
  \Bigg).
\end{multline} 
Taking into account frequency equation~\eqref{fr-eq-dim}, we get 
\begin{equation}
  J=
  \frac{{\mathcal C}^2}{2}\left( 
    \frac{\Omega_0
      \big(
        3\rho B^{1/4}+\sqrt2M\sqrt[4]{k-\rho\Omega_0^2}
      \big)
  }{4\sqrt2\sqrt[4]{k-\rho\Omega_0^2}} \right).
	\label{inv-1}
\end{equation}

Assuming now that $J$ is an adiabatic invariant, 
we obtain the following formula
\begin{equation}
 J=\const
 \quad\Longrightarrow\quad
|{\mathcal C}|\propto
  \frac{\sqrt[8]{k-\rho\Omega_0^2}}
	{\sqrt{\Omega_0(\sqrt{2}M\sqrt[4]{k-\rho\Omega_0^2}+3\rho\sqrt[4]{B})}},
  \label{propto}
\end{equation}
which is in agreement with Eq.~\eqref{amplitude}.

\section{Equivalent Hamiltonian system}
\label{sect-ham}
For the problem under consideration, the equivalent Hamiltonian system is \cite{arXiv.2602.18815}
\begin{equation}
  \dot{\mathcal P}=-\pd{\mathcal H}{\mathcal Q}
  ,
  \qquad
  \dot{\mathcal Q}=\pd{\mathcal H}{\mathcal P},
  \label{ham-sys}
\end{equation}
where
\begin{equation}
  \mathcal H=\mathcal H(\mathcal Q,\mathcal P,t)\=\frac{\mathcal P^2}{2\mathcal M(\et)}+\frac {\mathcal K(\et)\mathcal Q^2}2
  -\mathcal Qp(t)
  \label{ham}
\end{equation}
is the corresponding Hamiltonian, 
\begin{equation}
\mathcal Q=w(0,t),\qquad\mathcal P=\mathcal M\frac {\d w(0,t)}{\d t}
\label{QP}
\end{equation}
are the generalized
co-ordinate and the generalized momentum, respectively. The coefficients 
\begin{gather} 
  \mathcal M(\et)=\frac1{\Omega_0(\epsilon t) |C_0(\epsilon t)|}
  ,
  \label{M-eff}
  \\
  \mathcal K(\epsilon t)= \mathcal M(\epsilon t)\Omega_0^2(\epsilon t)
  \label{freq-harmonic}
\end{gather} 
are the effective mass and the effective stiffness, respectively. Quantity $|C_0(0)|$ is defined as the amplitude of 
free undamped oscillation of the discrete inclusion subjected to the impulse
force $\delta(t)$ for the system with constant parameters \eqref{all-slow0},
i.e., by Eq.~\eqref{C00}. Quantity $|C_0(\epsilon t)|$ is
defined by the same formula Eq.~\eqref{C00}, where the system parameters vary
according to Eq.~\eqref{all-slow-et}.

The evolution of the amplitude $\mathcal U$ can be found by the WKB approach.
One has \cite{arXiv.2602.18815}, Appendix~B:
\begin{gather}
\mathcal Q(t) =
|\mathcal C(\epsilon t)|
\cos\big(\psi- \arg \mathcal C\big)+\text{o}(1), 
\label{sol-eg0-eff}
\\
|\mathcal C|=
|\mathcal C(\epsilon t)|=|C(0)|{\frac{|\mathcal C_0(\epsilon t)|}{|\mathcal C_0(0)|}},
\label{amp-eff}
\\
\arg \mathcal C=\arg C(0),\qquad
\arg \mathcal C_0=\arg C_0(0),
\label{phase1}
\\
|\mathcal C_0|=
|\mathcal C_0(\epsilon t)|\=
\frac1{\sqrt{\mathcal M\Omega_0}},
\label{amp-expl-eff}
	\\
  |C(0)|=\big|C_0(0)\, p_F(\Omega_0(0))\big|,
\\
\arg C(0)=\arg p_F\big({\O_{0}(0)}\big)
+
\arg C_0(0)
.
\end{gather}
In Eq.~\eqref{sol-eg0-eff},
$\psi$ is defined by Eq.~\eqref{psi-def}; 
symbol $p_F\big({\O_{0}(0)}\big)$ is explained by Eq.~\eqref{pF}.

On the other hand, one can see that 
Eqs.~\eqref{amp-eff}--%
\eqref{amp-expl-eff}
are formally equivalent to
Eqs.~\eqref{amp}--\eqref{phase}, \eqref{amplitude}, respectively, and, therefore, again lead to Eq.~\eqref{propto}.
Moreover, Eq.~\eqref{amp-expl-eff}
indicates that
  \begin{equation}
    |\mathcal C_0(\epsilon t)|=\sqrt{|C_0(\epsilon t)|},
    \label{C0C0}
  \end{equation}
see Eqs.~\eqref{amplitude}, \eqref{C00}.
The analogous result was previously obtained
\cite{Gavrilov2024nody,arXiv.2602.18815}
in the framework of the problem
concerning localized oscillation of the string on the Winkler foundation.

\section{Numerics}
\label{sect-num}
To obtain numerical results we consider the problem for a sufficiently long
beam described by Eq.~\eqref{eq-semi} for $0<x<L$ clamped at $x=L$. 
The
non-dissipative case 
 \eqref{non-diss}
is under consideration.

\begin{figure}[h] 
\centering{\includegraphics[width=0.7\textwidth]{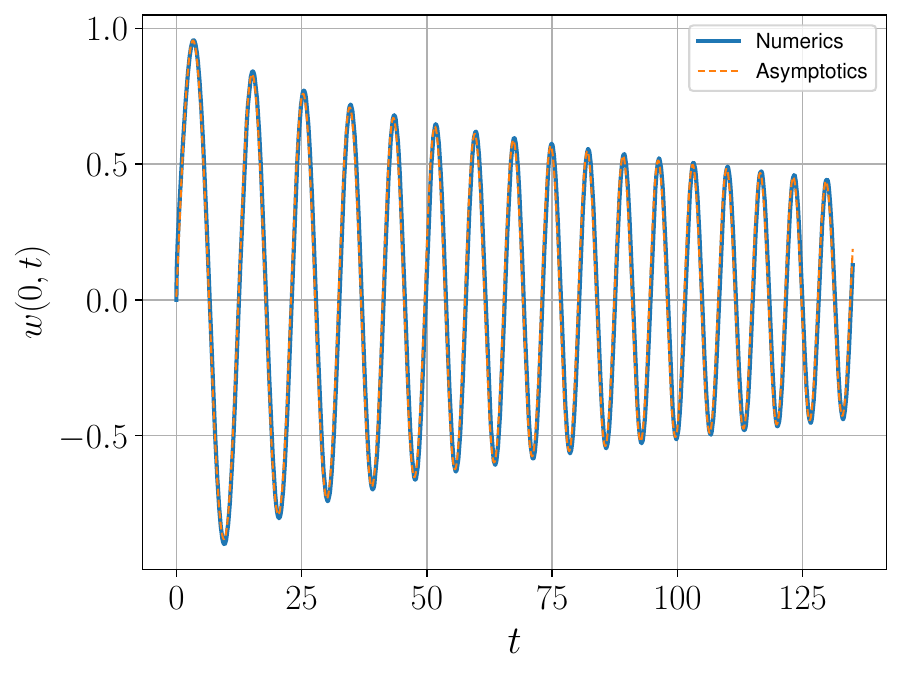}}
\caption{
The displacement $w(0,t)$ for the case when parameters
are taken according to Eq.~\eqref{non-diss}, \eqref{all-time1}, \eqref{all-time2}}
\label{u-time.pdf}
\end{figure}

To discretize PDE~\eqref{eq-semi}
in the particular case under investigation,
we use the following scheme
\cite{strikwerda2004finite}:
\begin{equation}
\label{eq-numeric}
\frac{{w}_j^{i+1}-2{w}_j^{i}+{w}_j^{i-1}}{(\Delta t)^2}
+
\frac{{w}_{j+2}^{i}-4{w}_{j+1}^{i}+6{w}_j^{i}-4{w}_{j-1}^{i}+{w}_{j-2}^{i}}{(\Delta x)^4}
+
\frac{u_j^{i+1}+u_j^{i-1}}2=0,
\end{equation}
where integers $i,\ j$ ($0\leq j\leq N,\ -1\leq i$) are such that
\begin{gather}
{w}_j^i={w}(j\Delta x,i\Delta t)
.
\end{gather}
Numerical boundary conditions corresponding to Eqs.~\eqref{angle-half-beam-dim},
\eqref{bc3-semi} are
\begin{equation}
  w_0^{i+1}-w_1^{i+1}+w_0^{i-1}-w_1^{i-1}=0,
\end{equation}
\begin{multline}
\label{bc-numeric}
\frac{-{w}_0^{i+1}+3{w}_1^{i+1}-3{w}_2^{i+1}+{w}_3^{i+1}}{(\Delta x)^3}
+
\frac{-{w}_0^{i-1}+3{w}_1^{i-1}-3{w}_2^{i-1}+{w}_3^{i-1}}{(\Delta x)^3}
\\=
\frac{K^{i+1}{w}_0^{i+1}+K^{i-1}{w}_0^{i-1}}2
+
\frac{p^{i+1}+p^{i-1}}2
-
M^i\,\frac{{w}_0^{i+1}-2{w}_0^{i}+{w}_0^{i-1}}{(\Delta t)^2}
-\frac{(M^{i+1}-M^{i-1})(w_0^{i+1}-w_0^{i-1})}{4(\Delta t)^2}
,
\end{multline}
where
\begin{equation}
K^i=K(i\Delta t),
\qquad
M^i=M(i\Delta t).
\end{equation}
At the right end, we use boundary conditions
\begin{gather}
w^i_{N}=w^i_{N-1}=0,
\end{gather}
{which correspond to the physical boundary conditions at $x=L$ in the form of
  \begin{gather}
    w=0,
    \\
    w'=0.
  \end{gather} 
}%
Numerical initial conditions are 
\begin{equation}
w^0_j=w^{-1}_j=0.
\label{num-ic}
\end{equation}
All numerical results below are obtained for the choice
\begin{gather}
\Delta t=2.5\cdot10^{-5}, \qquad 
\Delta x=1.1\sqrt{2\Delta t}\approx7.8\cdot 10^{-3}
\label{OSC-Deltas}
\end{gather}
to satisfy the stability condition \cite{strikwerda2004finite}:
\begin{equation}
  \frac {\Delta t}{(\Delta x)^2}< \frac12.
\end{equation}
The beam length is taken as 
\begin{equation}
 L=3t.
\end{equation}
The external force $p(\tau)$ for numerics is taken in the form of
\begin{equation}
p(\tau)=p_\ast\big(H(\tau)-H(\tau-\tau_0)\big),
\label{finite-step}
\end{equation}
wherein $p_\ast=1/\tau_0$. Thus, $p(\tau)\to\delta(\tau)$ as $\tau_0\to0$ in the weak
sense. The corresponding asymptotics is taken for the case
$p(\tau)=\delta(\tau)$.
In all following examples, the value of the small parameter is 
\begin{equation}
\epsilon=0.01,
\end{equation}
and 
\begin{equation}
\tau_0=0.01.
\end{equation}
The time-varying parameters are $M$ and $K$:
\begin{gather}
    M(T) = T,
\label{all-time1}
\\
    K(T)=-2.4+2.5T.
\label{all-time2}
\end{gather}%
In Fig.~\ref{u-time.pdf} we compare the corresponding asymptotic and numerical
solutions.
One can see that the solutions
are in  excellent agreement.

\section{Conclusion}
\label{sect-concl}
The main results of the paper 
is Eqs.~\eqref{sol-dim}--\eqref{sol-dim-D}, which
describe the amplitude evolution of localized oscillations of an Euler–Bernoulli
beam resting on a visco-elastic foundation and coupled with a discrete
oscillator. These formulas were derived via the asymptotic approach in
Sect.~\ref{sect-asy}. In the particular case of the non-dissipative system
\eqref{non-diss}, they were independently derived using two alternative
simplified analytical approaches in Sects.~\ref{sect-invariant}
and~\ref{sect-ham} as Eqs.~\eqref{propto} and~\eqref{amp-expl-eff},
respectively. Finally, the analytical predictions were validated numerically
in Sect.~\ref{sect-num}, demonstrating excellent agreement with the numerical
solutions.

Furthermore, we demonstrated that the method based on the equivalent
Hamiltonian system proposed in \cite{arXiv.2602.18815} for a string remains
applicable to the beam problem. Under this approach, determining the evolution
law for the amplitude $\mathcal C_0(\et)$ in the system with time-varying
coefficients \eqref{all-slow-et}, \eqref{non-diss} requires only knowledge of
the amplitude of inclusion oscillations
under delta-like excitation in the
corresponding system with constant coefficients \eqref{all-slow0}.
When solving the first-order approximation equation \eqref{od-eq-dim},
anticipating that relation~\eqref{C0C0} holds enabled us to deduce the
appropriate form of the solution. This confirms that the equivalent
Hamiltonian system method can be extended to complex systems where applying
the asymptotic approach directly is intractable. In particular, we expect
that this method will be useful for problems involving inertial loads
moving along beams, as well as in dynamic contact problems.



\appendix

\section{The dispersion relation}
In what follows, we consider the unperturbed  problem with 
$\epsilon=0$. Accordingly, all system parameters are constants
\begin{equation}
\begin{gathered}	
K=K(0),
\qquad
M=M(0),
\\
B=B(0),
\quad
\rho=\rho(0),
\quad
k=k(0),
\end{gathered}
\label{all-slow0}
\end{equation}
which are equal to the corresponding initial values at $t=0$. 

The dispersion relation for the operator in the left-hand side of Eq.~\eqref{eq}
is
\begin{equation}
	{B}\omega^4+\rho(\Omega_\ast^2-\Omega^2)=0,
	\label{disp}
\end{equation}
where $\omega$ is the wave-number, $\Omega$ is the frequency, $\Omega_\ast$ is defined by 
Eq.~\eqref{w-a-half-beam-dim}.
We deal with the right part of the beam $x>0$,
therefore, we consider the two roots of the dispersion relation given by the expression
\begin{equation}
	\omega(\Omega) = \omega^{\alpha}(\Omega) = a(\Omega)(\I+\alpha).
	\label{omega-delta}
\end{equation}
Here $\alpha=\pm 1$,
\begin{equation}
	a(\Omega)=\sqrt[4]{\frac{\rho(\Omega_\ast^2-\Omega^2)}{4 B}}.
  \label{a-def}
\end{equation}


\section{The Green function}
Here we deal with the semi-infinite ($x>0$) beam without mass-spring
inclusion in the case $p=\delta(t)$.
Consider the governing equation in the frequency domain:
\begin{equation}
 -\Omega^2\rho {G} + B G'''' + k G = 0
\label{FB-beam-Green-alt-maineq}
\end{equation}
with boundary conditions 
\begin{align}
  &G'\big|_{x=0}=0,
\label{FB-beam-bc1-s}
	\\
	&G'''\big|_{x=0}=\frac{1}{2B},
\label{FB-beam-bc3-s}
  \\
  &G\big|_{x\to+\infty}=0,
  \label{bcinf1}
  \\
  &G'\big|_{x\to+\infty}=0.
  \label{bcinf2}
\end{align}
Here $G$ is the Green function for a beam without inclusion in frequency domain.

We look for the solution of the formulated problem in the form of
\begin{gather}
G=
{\mathcal W}(\Omega)\exp(\I \omega^{1} x)+\bar{{\mathcal W}}(\Omega)\exp(\I\omega^{-1}x)=
{\mathcal W}(\Omega)\exp(\I \omega^{1} x)+\cc,
\label{FB-beam-spectrum-alt-r}
\end{gather}
where $\omega^{\alpha}(\Omega)$, $\alpha=\pm1$, is given by Eq.~\eqref{omega-delta},
$a(\Omega)$ should be found via Eq.~\eqref{a-def}.
The representation 
\eqref{FB-beam-spectrum-alt-r}
satisfies boundary conditions \eqref{bcinf1}, \eqref{bcinf2}.
The boundary conditions \eqref{FB-beam-bc1-s}, \eqref{FB-beam-bc3-s} can be rewritten as follows:
\begin{gather}
{\mathcal W}\omega^{1}+\bar{{\mathcal W}}\omega^{-1}=0,
\\
-\I({\mathcal W}(\omega^{1})^3+\bar{{\mathcal W}}(\omega^{-1})^3)=\frac{1}{2B}.
\end{gather}
Resolving these equations yields
\begin{equation}
{\mathcal W}=\frac{\I}{2B\omega^{1}((\omega^1)^2-(\omega^{-1})^2)}=\frac{1-\I}{16Ba^3}.	
\end{equation}
Taking into account Eq.~\eqref{a-def}, one has
\begin{equation}
{\mathcal W}=\frac{(1-\I)\sqrt{2}}{8B^{1/4}\rho^{3/4}(\Omega_\ast^2-\Omega^2)^{3/4}},
\end{equation}
and, finally,
\begin{equation}
G=\frac{\sqrt{2}}{8Ba^3}\exp(-a x)\cos\left(a x-\frac{\pi}{4}\right).
	\label{Gr-func-dim}
\end{equation}

\section{The spectral problem}
Here we deal with the semi-infinite ($x>0$) beam with the mass-spring
inclusion in the case $p=0$.
Consider the governing equation in frequency domain:
\begin{equation}
	-\Omega^2\rho{W} + B {W}{''''}+ k{W} = \big(M\Omega^2 {W}(0) - K {W}(0)\big)\delta(x).
	\label{eq:2.1-dim-amp}
\end{equation}
Here we look for the modes corresponding to the frequencies from the discrete part of the spectrum (the trapped modes), 
which lies below the cut-off frequency $\Omega_\ast$:
\begin{equation}
	0<\Omega<\Omega_\ast.
  \label{O-interval}
\end{equation}
The trapped modes are modes with finite energy, therefore we require \cite{Ind-book-R2E}
\begin{equation}
	\int_{0}^{+\infty}\,{{W}}^2 \,d\xi <\infty, \qquad 
  \int_{0}^{+\infty}\,({{W}''}){}^2 \,d\xi <\infty.
\end{equation}

Using Eq.~\eqref{Gr-func-dim}, one obtains the following expression for the amplitude:
\begin{equation}
	{W}(x)={W}(0)\frac{(M\Omega^2-K)}{8Ba^3}\exp{(-a x)}
	\cos\left(a x -\frac{\pi}{4}\right).
\end{equation}
Calculating the solution at $x=0$ and taking into account Eq.~\eqref{a-def}, one obtains:
\begin{equation}
	{W}(0)=\frac{(M\Omega^2-K)\sqrt{2}}{4 B^{1/4}\rho^{3/4}(\Omega_\ast-\Omega)^{3/4}}{W}(0).
\end{equation}
Thus, the frequency equation is 
\begin{equation}
	(\Omega_\ast^2-\Omega^2)^{3/4}=\frac{\sqrt{2}(M\Omega^2-K)}{4 B^{1/4}\rho^{3/4}}.
	\label{fr-eq-dim}
\end{equation}

One can show that the solution of the frequency equation
\eqref{fr-eq-dim} in interval \eqref{O-interval}
exists if and only if 
\begin{equation}
-\frac{4 B^{1/4} k^{3/2}}{\sqrt{2}\rho^{3/4}}	\leq K \leq \frac{Mk}{\rho},
	\label{K-restr}
\end{equation}
it is unique and such that
\begin{gather}
	0 \leq \Omega_0^2 \leq \frac{k}{\rho}, \quad K < 0,
	\\
	\frac{K}{M} \leq \Omega_0^2 \leq \frac{k}{\rho}, \quad K \geq 0.
\end{gather}
Inequality \eqref{K-restr} defines the localization domain \cite{Gavrilov2024nody} in the problem
parameters space.

The form of free oscillations at $\Omega=\Omega_0$  can be calculated as follows:
\begin{equation}
	w(x,t)= {\tilde C}G(x,\Omega)\exp(-\I\Omega_0 t)
    +\cc,
\end{equation}
where $\tilde {C}$ is an arbitrary complex constant. The above equation is equivalent to the following one:
\begin{equation}
	w(x,t)=\bar w(x,t)\=|C|\exp{(-a x)}
	\cos\left(a x-\frac{\pi}{4}\right)\cos(\Omega_0 t-\arg C),
  \label{free-osc}
\end{equation}
where
\begin{equation}
|C|=\dfrac{\sqrt{2}|\tilde {C}|}{4Ba^3}, 
\qquad
\arg {C}=\arg \tilde{C}.
\end{equation}

\section{Inhomogeneous non-stationary problem}
Consider the non-stationary problem with $p\neq0$.
Applying to 
Eq.~\eqref{eq} the Fourier transform in time $t$ results in 
\begin{equation}
	Bw_F'''' + (k-\rho \Omega^2)w_F  =  \big((M\Omega^2-K)w_F(0,\Omega)+p_F(\Omega)\big)\delta(x), 
	\label{ino-BEAM-non-dim}
\end{equation} 
where $w_F(x,\Omega),\ p_F(\Omega)$ are the Fourier transforms of
$w(x,t)$ and $p(t)$, respectively.
The solution of Eq.~\eqref{ino-BEAM-non-dim} is
\begin{equation}
	w_F(x,\Omega)
	=\int_0^{x}
	G(\xi, \Omega)
	\big((M\Omega^2-K)w_F(0,\Omega)+p_F(\Omega)\big)\delta(\xi)\mbox{d}\xi.
\end{equation}
Using expressions~\eqref{Gr-func-dim},  we have
\begin{equation}
	w_F(0,\Omega)=\frac{(M\Omega^2-K)w_F(0,\Omega)+p_F(\Omega)}{2\sqrt{2}B^{1/4}\rho^{3/4}(\Omega_\ast^2-\Omega^2)^{3/4}}.
\end{equation}
Expressing $w_F(0,\Omega)$ from the above equation we get
\begin{equation}
	w_F(0,\Omega)=\frac{p_F(\Omega)}{K-M\Omega^2 +2\sqrt{2}B^{1/4}\rho^{3/4}(\Omega_\ast^2-\Omega^2)^{3/4}}.
	\label{w_FF-dim}
\end{equation}
The inverse transform of $w_F(0,\Omega)$ yields
\begin{equation}
	w(0,\tau)
	=\frac1{2\pi}\Int
	\frac{p_F(\Omega)\exp^{-\mbox{i}\Omega \tau}\,\mbox{d}\Omega}{K-M\Omega^2 +2\sqrt{2}B^{1/4}\rho^{3/4}(\Omega_\ast^2-\Omega^2)^{3/4}}.
	\label{before-stat-phase-dim}
\end{equation}
Applying the residue theorem,  Jordan's lemma, and the method of stationary phase
to asymptotic
evaluation of the integral in
the right-hand side of \eqref{before-stat-phase-dim} {results in} \cite{Gavrilov1999jsv,Fedoryuk1977}
\begin{multline}
	w(0,\tau)=
	-\I \sum_{\bar{\Omega}=\pm\Omega_0-\i0}\!\!\!p_F(\bar{\Omega})
	\res \left(
	\frac{1}{K-M\Omega^2 +2\sqrt{2}B^{1/4}\rho^{3/4}(\Omega_\ast^2-\Omega^2)^{3/4}} , \bar{\Omega}\right)
	\exp (-\i\bar\Omega t) +o(1),  \\
	t\to\infty.
	\label{undamped-as-gengen-dim}
\end{multline}
Here symbol $\res\big(f(\Omega),\bar\Omega\big)$ means the residue of function
$f(\Omega)$ at a pole $\Omega=\bar\Omega$.
The terms $-\i0$ in the expression for the poles
\begin{equation}
	\bar\Omega=\pm\Omega_0-\i0
	\label{poles}
\end{equation}
are taken in accordance with the limiting absorption principle.
One has 
\begin{equation}
	\res\!\left(
	\frac{1}{K-M\Omega^2 +2\sqrt{2}B^{1/4}\rho^{3/4}(\Omega_\ast^2-\Omega^2)^{3/4}},\pm \Omega_0-\i 0\right)
		=\mp	
	\frac{\sqrt[4]{\Omega_{\ast}^2-\Omega_0^2}}{\Omega_0(2M\sqrt[4]{\Omega_{\ast}^2-\Omega_0^2}+3\sqrt{2}B^{1/4}\rho^{3/4})},
\end{equation}
thus,
\begin{gather}
	w(0,t)=|C(0)|
    \cos\big(\Omega_0(0) t-\arg C(0)\big)
    +o(1),\quad t \to\infty,
	\label{undamped-as-general-pnotzero-dim}
	\\
  |C(0)|=\big|C_0(0)\, p_F(\Omega_0(0))\big|,
  \label{C00-mod}
  \\ 
  |C_0(0)|=\left.\frac{\sqrt2\sqrt[4]{k-\rho\Omega_0^2} 
	}
  {\Omega_0(3\rho\sqrt[4] B+\sqrt2M\sqrt[4]{k-\rho\Omega_0^2})}
  \right|_{T=0},
  \label{C00}
\\
\arg C(0)=\arg p_F\big({\O_{0}(0)}\big)
+
\arg C_0(0)
,
\label{argC-eff}
\\
\arg C_0(0)=
\frac\pi2
.
\label{Co-value-eff}
\end{gather}
Here, the symbol $p_F\big(\O_0(0)\big)$ denotes the Fourier transform of $p(t)$:
\begin{equation}
p_F=\int_{-\infty}^{+\infty}p(t){\mathrm e}^{\I\Omega t}\,\d t
\label{pF}
\end{equation}
calculated
at $\Omega=\Omega_0(0)$. 
Note that $C(0)=C_0(0)$ in the case $p=\delta(t)$.
Hence, for large times, the
non-stationary response of the system under consideration is undamped
oscillations with the trapped mode frequency $\Omega_0$.

\bibliographystyle{plainnat.bst}
\bibliography{bib/serge-gost,bib/all}

\selectlanguage{english}